\documentclass[a4paper,11pt]{article}
\pdfoutput=1 

\usepackage{jcappub} 

\usepackage{bm}
\usepackage{url}
\usepackage{amsmath}
\usepackage{amsfonts}
\usepackage{graphicx}
\usepackage{color}
\usepackage{hyperref}
\usepackage{multirow}

\newcommand{\be}{\begin{equation}}
\newcommand{\ee}{\end{equation}}
\newcommand{\bea}{\begin{eqnarray}}
\newcommand{\eea}{\end{eqnarray}}

\newcommand{\lsim}{\mathrel{\mathop{\kern 0pt \rlap
  {\raise.2ex\hbox{$<$}}}
  \lower.9ex\hbox{\kern-.190em $\sim$}}}
\newcommand{\gsim}{\mathrel{\mathop{\kern 0pt \rlap
  {\raise.2ex\hbox{$>$}}}
  \lower.9ex\hbox{\kern-.190em $\sim$}}}

\newcommand{\GeV}{{\rm GeV}}

\newcommand{\TeV}{{\rm TeV}}

\newcommand{\m}{{\rm m}}

\newcommand{\s}{{\rm s}}
\newcommand{\sr}{{\rm sr}}

\begin{document}

\title{Model independent interpretation of recent CR lepton data after AMS-02}

\author[a,b]{Daniele Gaggero}
\affiliation[a]{SISSA, via Bonomea 265, 34136 Trieste (Italy)}
\affiliation[b]{INFN, Sezione di Trieste, via Valerio 2, 34127 Trieste (Italy)}

\author[c,d]{Luca Maccione}
\affiliation[c]{Arnold Sommerfeld Center, Ludwig-Maximilians-Universit\"{a}t, Theresienstra{\ss}e 37, D-80333 M\"{u}nchen, Germany}
\affiliation[d]{Max-Planck-Institut f\"{u}r Physik (Werner Heisenberg Institut), F\"{o}hringer Ring 6, D-80805 M\"{u}nchen, Germany}

\emailAdd{dgaggero@sissa.it}
\emailAdd{luca.maccione@lmu.de}

\subheader{LMU-ASC 43/13; MPP-2013-161}

\date{\today}

\abstract{
We model the CR leptonic fluxes above 20 GeV in terms of a superposition of a standard and a charge symmetric extra component, which we generically describe as power-laws in momentum. We investigate under these hypotheses the compatibility between AMS-02, Fermi-LAT and PAMELA datasets on positron fraction, electron+positron spectrum and electron spectrum respectively. We find that it is possible to reconcile AMS and Fermi-LAT data within  uncertainties, if energy-dependent effects are present in Fermi-LAT systematics. We also study possible deviations from charge symmetry in the extra component and find no compelling evidence for them.
}

\maketitle
\flushbottom

\section{Introduction}

The AMS collaboration recently published a very accurate measurement of the positron fraction (PF) in cosmic rays (CR) from 0.5 to 350 GeV \cite{Aguilar:2013qda}. 
The data show the presence of a clear rise above 10 GeV and confirm former experimental results by PAMELA { in the energy range 1.5--100~GeV \cite{Adriani:2008zr} and Fermi-LAT in the energy range 20--200~GeV \cite{FermiLAT:2011ab}.} The AMS collaboration also extended the energy range over which the PF was measured. As a result, the rise of the PF continues up to 250 GeV, while above that energy a precise evaluation of the spectral slope is not possible because of poor statistics. { Albeit confirming the increase above 20~GeV, the PF measured by AMS shows a few differences with respect to previous measurements by PAMELA and Fermi-LAT, namely a lower normalization with respect to Fermi-LAT, and a gentler increase above 50~GeV with respect to PAMELA.}

Soon after publication of the data, the issue of the mutual compatibility of these datasets was raised in several papers \cite{Yuan:2013eba,Yuan:2013eja,Jin:2013nta,Cholis:2013psa}. Analyses were typically performed within phenomenological scenarios in which a ``standard'' leptonic component originating from acceleration in Supernova Remnants (SNRs) and secondary production in the interstellar medium is added to an extra component with a harder spectrum. The nature of this extra component is still unknown and may be either of astrophysical origin \cite{Hooper:2008kg,Blasi:2009hv,Mertsch:2009ph,Ahlers:2009ae}  or of exotic origin (see \cite{Cholis:2013psa,Linden:2013mqa} and Refs. therein). 

In \cite{Yuan:2013eja,Cholis:2013psa} it was shown that, with conventional choices of the standard component and a contribution from DM particles annihilating into leptonic species it is not possible to simultaneously reproduce the AMS PF and Fermi-LAT measurements \cite{Abdo:2009zk,Ackermann:2010ij} of the electron+positron (CRE) spectrum. However, models in which DM particles annihilate into light intermediate states which then decay into a combination of muons and charged pions are compatible with those data, provided a { break} in the primary electron component is assumed, { with a harder slope above the break}. Such hardening may be explained considering the contribution from a local source that accelerates only electrons (e.g.~a SNR). 
On the other hand, Yuan et al. \cite{Yuan:2013eba} use a Markov Chain MonteCarlo (MCMC) sampler to fit the datasets considering both an astrophysical scenario in which pulsars emit electron+positron pairs and a DM scenario with leptophilic annihilation/decay channels. They also find a tension between AMS PF and Fermi-LAT data, and claim that AMS-02 data require less contribution from the extra component with respect to Fermi-LAT. 


These recent results may be understood if a charge asymmetry is present in the extra source \cite{Cholis:2013psa,Yuan:2013eba,Masina:2013yea}. We remark that this might be expected in both exotic and astrophysical scenarios. For example, if DM decays by violating lepton flavor symmetry, it is natural to expect the spectra of the resulting electrons and positrons to be different, both in normalization and in energy dependence (see e.g.~\cite{Frandsen:2010mr,Masina:2011hu,Masina:2013yea} and Refs.~therein). If instead the extra source is of astrophysical origin, one possibility is that it includes the contributions of local pulsars and SNRs, not normally accounted for in galactic propagation codes, which would yield charge symmetric and asymmetric contributions, respectively. 

{ If confirmed,} the existence of a charge asymmetry would constitute a significant modification of the standard charge symmetric extra-component paradigm studied so far. 
Luckily enough, the issue is related to high-energy electrons and positrons. If we assume, as in the standard acceleration theory, that CREs are injected in the interstellar medium with a power-law energy spectrum (given that CREs are ultra-relativistic in this energy range, we will { identify} momentum with energy), and if we consider that above a few tens GeV { the shape of the observed spectrum only depends on the combined effects of energy losses and rigidity-dependent diffusion}, then we can approximate the propagated spectra with power-laws without losing generality. We can therefore apply standard data analysis techniques to study the properties of the observed spectra in a model independent way. 

This is the purpose of the present paper, that is composed as follows. In Section \ref{sec:method} we will discuss our methodology, including the assumptions we make and the numerical tools we use in our study. In Section \ref{sec:results} we will show the results of our investigation. Finally, in Section \ref{sec:conclusions} we will draw our conclusions.

\section{Method}
\label{sec:method}

\begin{figure}[tbp]
\begin{center}
\includegraphics[width=\textwidth]{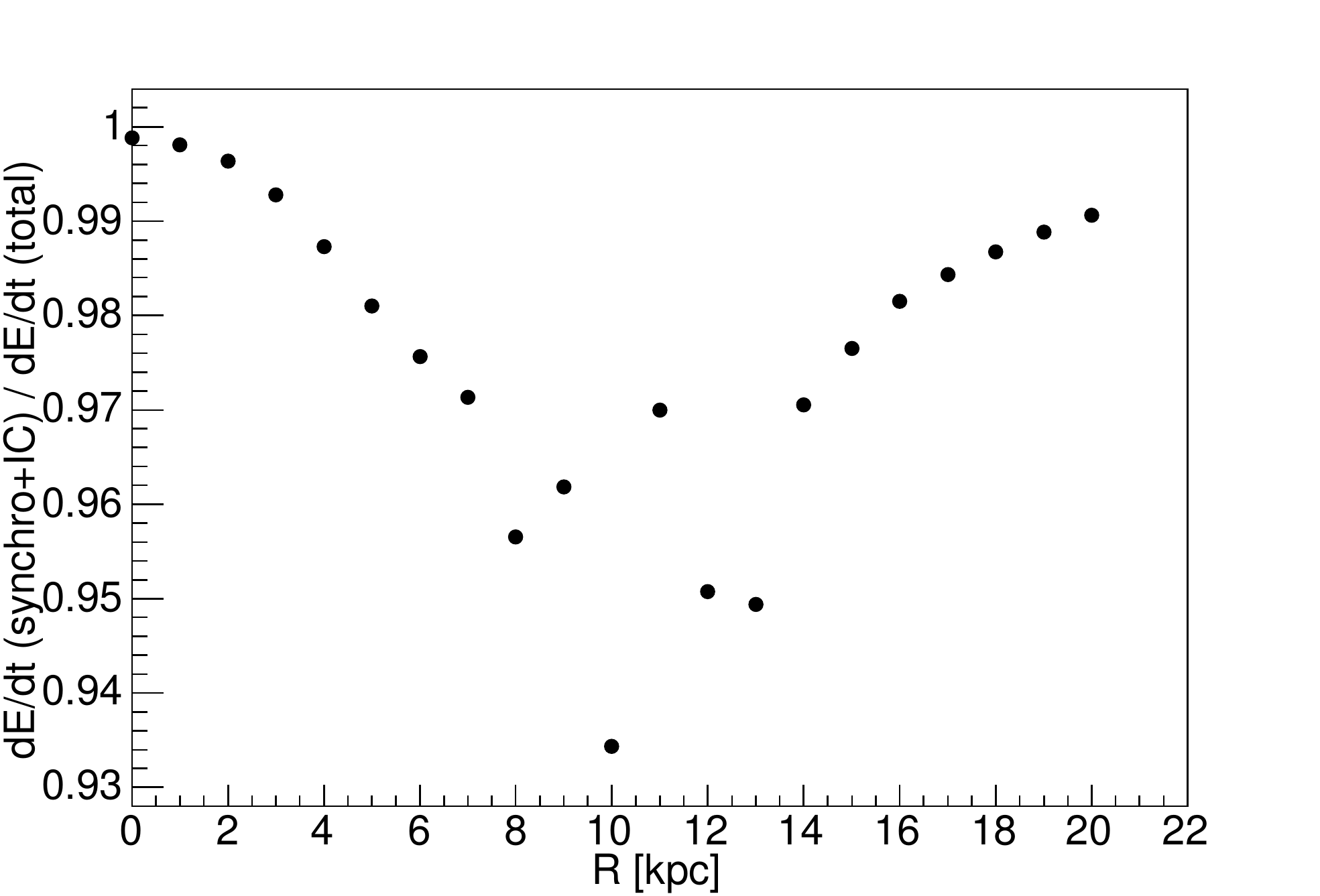}
\caption{Radial dependence of the ratio of energy loss rates for synchrotron and inverse Compton process with respect to the total energy loss rate on the galactic plane for $\sim100~\GeV$ electrons.}
\label{fig:eloss}
\end{center}
\end{figure}

We want to analyze the compatibility of the following 3 data sets in the high energy range:
%
a) Fermi-LAT \cite{Abdo:2009zk,Ackermann:2010ij} and HESS \cite{Aharonian:2009ah} spectra of $e^+ + e^-$;
b) PAMELA spectrum of $e^-$ only \cite{Adriani:2011xv};
c) AMS positron ratio $e^+/(e^+ + e^-)$ \cite{Aguilar:2013qda}.
%
Above 20 GeV the effects of solar modulation, re-acceleration and convection are negligible. Moreover, the contribution of secondary electrons and positrons is negligible at these energies. Therefore, the propagated spectrum can only be affected by \cite{Berezinsky_book}: 
\begin{itemize}
\item The energy dependent escape from the galaxy, described by the diffusion coefficient, typically assumed as $D=D_{0}(\rho/\rho_{0})^{\delta}$, where $\rho$ is the particle rigidity and $\delta$ is a constant parameter;
\item Continuous energy losses, including ionization, coulomb scattering and bremsstrahlung in the interstellar gas, synchrotron losses in the galactic magnetic fields and inverse Compton scattering off interstellar radiation fields. However, above 100~GeV synchrotron and inverse Compton losses are the dominant processes, as we show in Fig.~\ref{fig:eloss} for the gas, magnetic and radiation field models typically used in Galprop \cite{Galpropweb,Strong:2007nh} and DRAGON \cite{Dragonweb,DiBernardo:2009ku,DiBernardo:2010is}. At 20~GeV the maximal contribution of the other processes is $\sim25\%$ and it decreases with increasing energy, becoming of the order of a few \% at 100~GeV. We remark that for both synchrotron and inverse Compton processes $dE/dt \propto E^{2}$ in the ultra-relativistic Thomson regime.
\end{itemize}

In these conditions, it is well known \cite{Bulanov_1974Ap&SS} that if a power-law energy (or momentum) spectrum is injected by CRE sources, then a power-law spectrum, with a possibly different slope and with breaks, is to be expected after propagation. In fact, we checked that in the energy range we are considering a pure power-law is expected after propagation. We can then assume with confidence, for our purposes, that the propagated spectra be perfect power-laws, plus some exponential cutoff at high energy, describing a possible cutoff at injection or the cutoff induced by the energy loss length becoming shorter than the typical distance between the Earth and the closest CRE sources. We start our analysis by assuming the presence of 2 components:
\begin{description}
\item[Component $\bm A$] a primary electron component with spectrum $J_A = A\times (E/E_{0})^{-\alpha} \exp(-E/E_{{\rm cut,}A})$
\item[Component $\bm B$] an extra component of electrons and positron with spectrum $J_{B}^{e^{-}} = J_{B}^{e^{+}} = J_B = B\times (E/E_{0})^{-\beta} \exp(-E/E_{{\rm cut,}B})$,
\end{description}
where $E_{0}=20~\GeV$ and the units of the parameters will be $\GeV^{-1}\m^{-2}\sr^{-1}\s^{-1}$ for $A$ and $B$, $\TeV$ for $E_{\rm cut,A}$ and $E_{\rm cut,B}$, while $\alpha$ and $\beta$ are numbers.
In this model-independent parametrization: 
%
a) the Fermi-LAT and HESS spectra are described by $(J_A \,+\, 2\,J_B)$, 
b) the PAMELA electron spectrum is $(J_A + J_B)$, and 
c) the AMS PF is $J_B/(J_A\,+\,2\,J_B)$.

Given that it is irrelevant in the analysis, because the primary electron components falls off very steeply with increasing energy, we fix the value of $E_{{\rm cut,}A}=10~\TeV$. With this choice, we are left with 5 free parameters. We use therefore a Markov Chain MonteCarlo (MCMC) algorithm to estimate the probability density functions of the parameters. We { employ} the Bayesian Analysis Toolkit (BAT) \cite{BATweb,2009CoPhC.180.2197C} to perform the MonteCarlo scan and the analysis. 

This setup is valid in the case of a charge symmetric extra-component. In order to study also { possible departures from the charge symmetric scenario}, we will introduce two possibilities. First, we will consider the case in which the electron and positron spectra of the extra component differ only in normalization. We will then introduce a parameter $\epsilon$ such that electrons correspond to $(1+\epsilon)J_{B}$ and positrons to $(1-\epsilon)J_{B}$. { Finally,} we will also consider the presence of a pure electron component $C$, { with} $J_{C} = C (E/E_{0})^{-\gamma}\exp(-E/E_{\rm cut,C})$, such that the electron and positron spectra of the total extra-component (B+C) may differ also in slope. { However, as we will discuss later on, the C component fit leads only to very small numerical adjustments that can be considered a numerical artifact.}

As a final remark, we note that with our formalism we cannot capture any irregular behavior of the extra component, as expected if several individual local sources contribute to it \cite{Malyshev:2009tw}. However, neither Fermi-LAT nor AMS-02 data show hints of this behavior, therefore we will neglect this possibility in the following.

\section{Results}
\label{sec:results}

We first consider our datasets separately, then we perform a combined analysis. {
 We assume a flat prior distribution for all the parameters. The parameter ranges are shown in Tab.~\ref{tab:parranges}.} 

\begin{table}[tdp]
\caption{Range of parameters sampled in our analysis. Flat priors are always assumed. Units of dimensional parameters are given in text.}
\begin{center}
\begin{tabular}{|c|c|c|c|c|c|}
\hline $A$ & $\alpha$ & $B$ & $\beta$ & $E_{\rm cut, B}$ & $\epsilon$  \\
\hline $130-150$ & $3.0-3.3$ & $8-12$ & $2.5-2.8$ & $1-4$ & $0-1$ \\
\hline
\end{tabular}
\end{center}
\label{tab:parranges}
\end{table}%

\subsection{Analysis of single datasets}
\label{subsec:single}

\begin{figure}[tbp]
\begin{center}
\includegraphics[width=\textwidth]{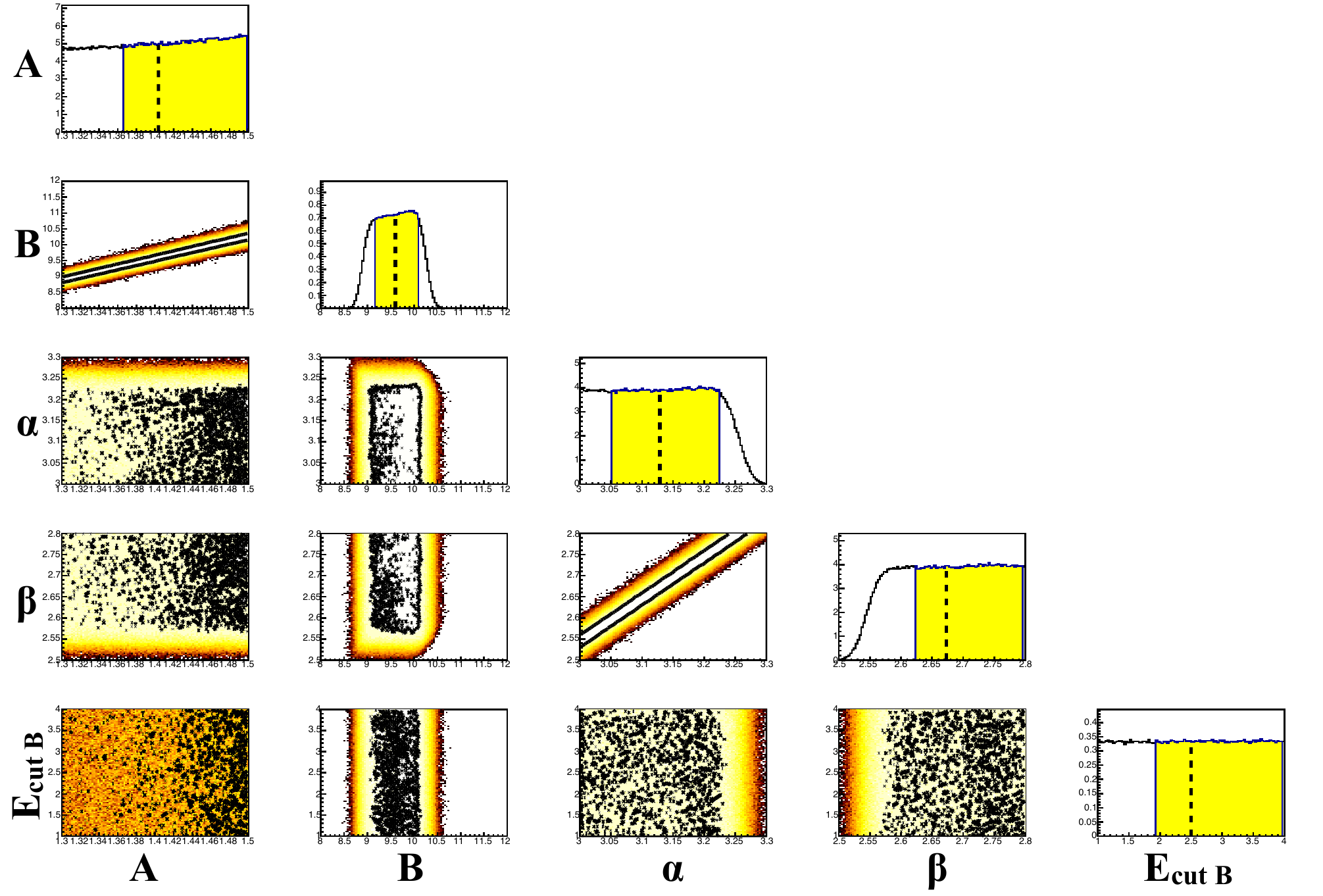}
\caption{{ Posterior probability distribution and correlation plots for the model parameters using AMS-02 data. The plots should be read in matrix form, where the quantities shown in a subplot are identified by the labels displayed in the corresponding row or column. A $ii$-type plot (the ones on the diagonal) represents the posterior probability distribution of a single parameter, with the yellow region highlighting the 68\% confidence area and the vertical black dashed line the median value. A $ij$-type plot shows instead the correlation between parameters $i$ and $j$. To improve readability we show the values of the parameter $A$ (first column) divided by 100. Dots and small islands in these plots are signal of noise in the MCMC algorithm, due to the poor sensitivity of data to the related parameters.}}
\label{fig:AMSparams}
\end{center}
\end{figure}

\begin{figure}[tbp]
\begin{center}
\includegraphics[width=\textwidth]{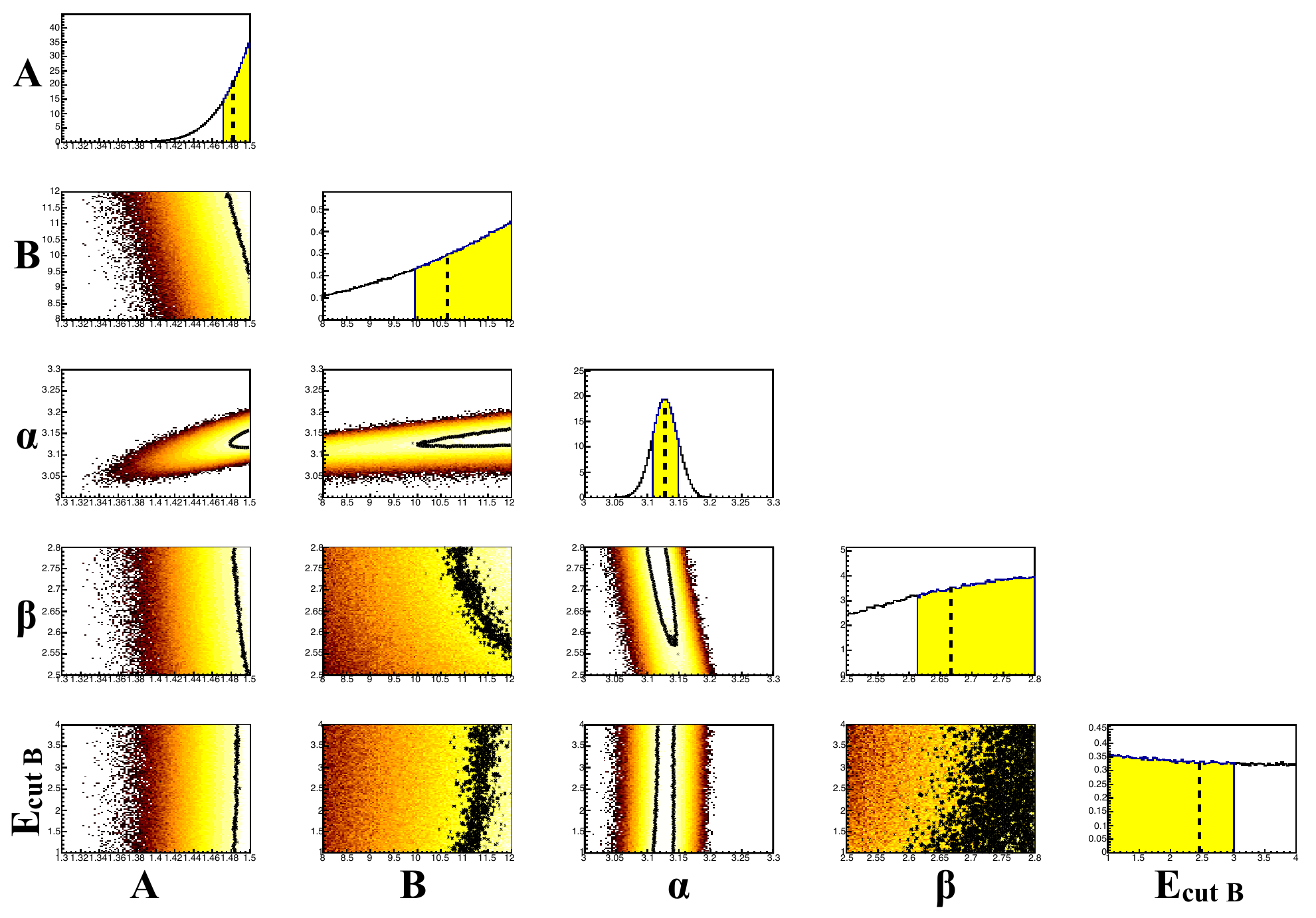}
\caption{Same format as Fig.~\ref{fig:AMSparams} for Pamela data.}
\label{fig:Pamelaparams}
\end{center}
\end{figure}

\begin{figure}[tbp]
\begin{center}
\includegraphics[width=\textwidth]{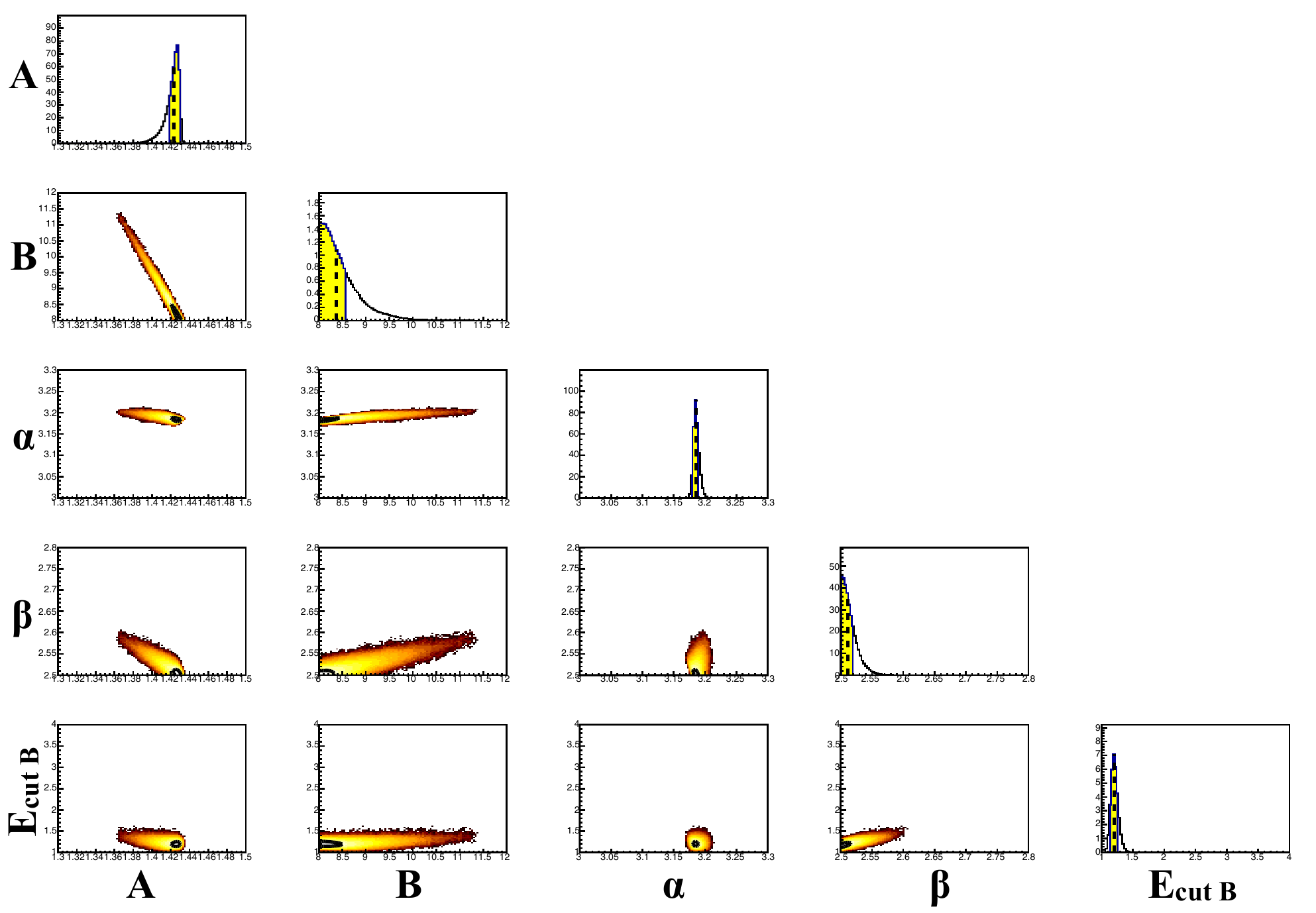}
\caption{Same format as Fig.~\ref{fig:AMSparams} for Fermi-LAT data.}
\label{fig:Fermiparams}
\end{center}
\end{figure}

We show in Figures \ref{fig:AMSparams}, \ref{fig:Pamelaparams} and \ref{fig:Fermiparams} the results of the analysis of the individual datasets of AMS-02, PAMELA and Fermi-LAT respectively. The AMS analysis (Fig.~\ref{fig:AMSparams}) shows a high degeneracy between most parameters, and does not point towards a well defined region in the parameter space. Of particular interest are the correlation plots for $A$-$B$ and $\alpha$-$\beta$. The two couples of parameters are very tightly correlated, indicating that a high degeneracy exists between the two components $A$ and $B$. This is not unexpected for the positron ratio data. Indeed, no matter how steep the primary component $A$ is taken, it is always possible to find a corresponding index for the component $B$ that produces a good fit for the ratio. Moreover, the posterior probability distribution for most of the parameters is flat, indicating that the positron fraction alone cannot disentangle the contributions of the two components.

A similar situation would hold for the PAMELA only analysis (Fig.~\ref{fig:Pamelaparams}) if we would only account for the positron fraction data. Since instead we include in this analysis only the electron spectrum measured by PAMELA, which is described very well by a single power-law, the degeneracy $\alpha$-$\beta$ is broken. In order to explain PAMELA electron data we do not need an extra component, and simply fitting the parameters of component $A$ would be enough. 

The Fermi-LAT analysis, instead, isolates a particular region in the parameter space (Fig.~\ref{fig:Fermiparams}).
By looking at the $\alpha$-$\beta$ correlation plot we can notice that a very narrow range of values for the two injection indexes is able to produce the observed slope of the electron+positron spectrum. However, while the preferred value of $\alpha$ is very well compatible, within uncertainties, with the value preferred by PAMELA, Fermi-LAT data point to very small values of $\beta$, which are about 2$\sigma$ away from the values preferred by PAMELA and also by AMS-02. This fact might already hint at the presence of some mild tension between the datasets. We also notice that Fermi-LAT data are able to constrain the cutoff energy of the $B$ component rather strongly, to $E_{\rm cut,B}\simeq 1.3~\TeV$.

\subsection{Combined analysis}
\label{subsec:combined}

\begin{figure}[tbp]
\begin{center}
\includegraphics[width=\textwidth]{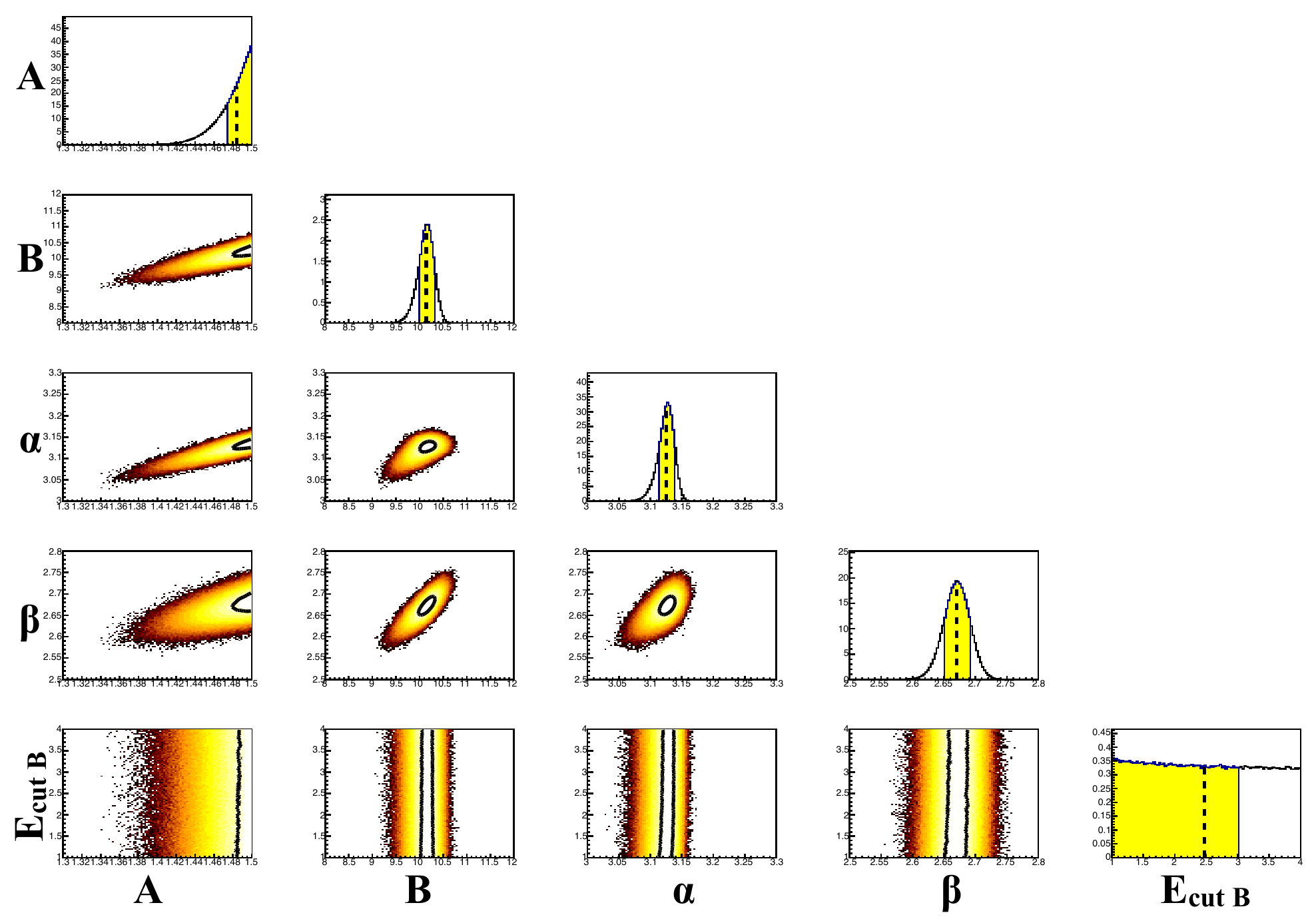}
\caption{Same format as Fig.~\ref{fig:AMSparams} for the combined AMS-02 and PAMELA data.}
\label{fig:PamelaAMSparams}
\end{center}
\end{figure}

Let us turn now our attention to the combined AMS and PAMELA analysis. From Fig.~\ref{fig:PamelaAMSparams} it is clear that the confidence regions for the two experiments nicely overlap. Moreover, by combining the electron spectrum of PAMELA with the high accuracy data of AMS-02 we can better pin-point the slope $\beta$ of the $B$ component. 

\begin{figure}[tbp]
\begin{center}
\includegraphics[width=\textwidth]{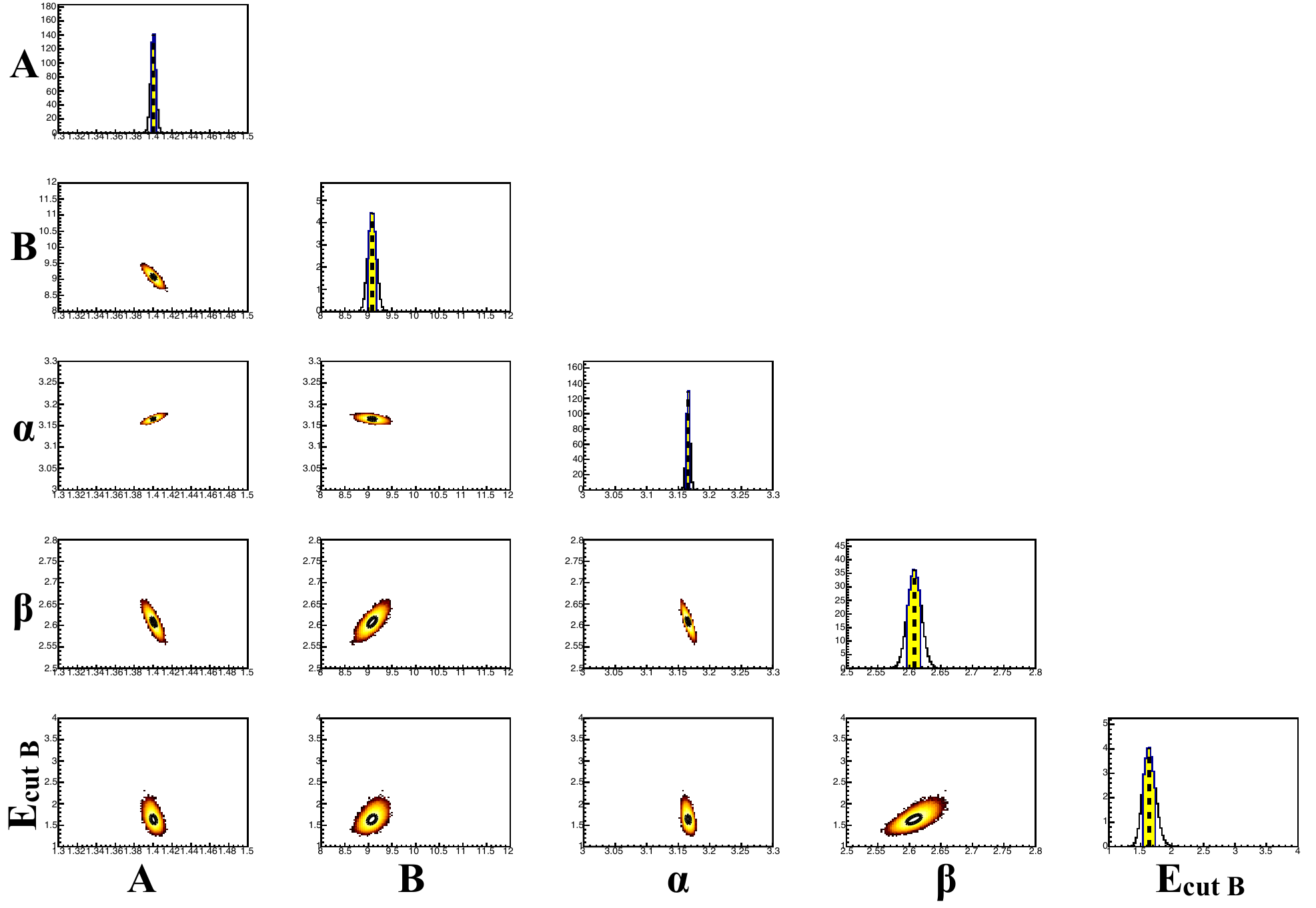}
\caption{Same format as Fig.~\ref{fig:AMSparams} for the combined AMS-02 and Fermi-LAT data.}
\label{fig:FermiAMSparams}
\end{center}
\end{figure}

\begin{figure}[tbp]
\begin{center}
\includegraphics[width=\textwidth]{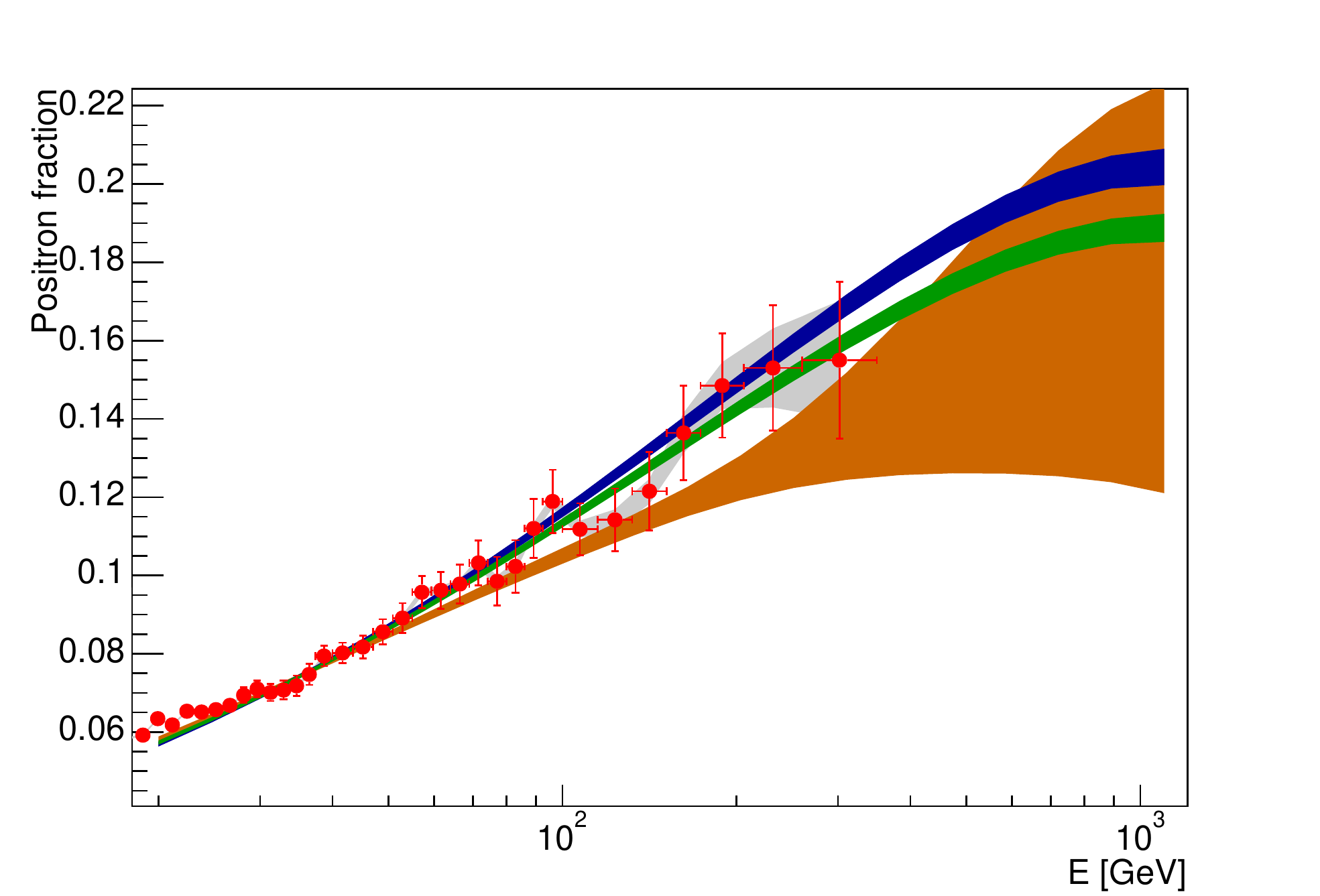}
\caption{Positron fraction for the best fit model derived combining AMS-02 and Fermi-LAT data. Red points are AMS-02 experimental data. The blue area represents our best-fit model in the charge-symmetric scenario, with its 68\% uncertainty band. The green area is instead for the best fit model assuming charge asymmetry in the $\epsilon$-parameterization. The orange area is finally for the best fit model assuming charge asymmetry with the C component. Error bars are statistical errors, while systematic errors correspond to the grey band.}
\label{fig:bestfitpf}
\end{center}
\end{figure}

\begin{figure}[tbp]
\begin{center}
\includegraphics[width=\textwidth]{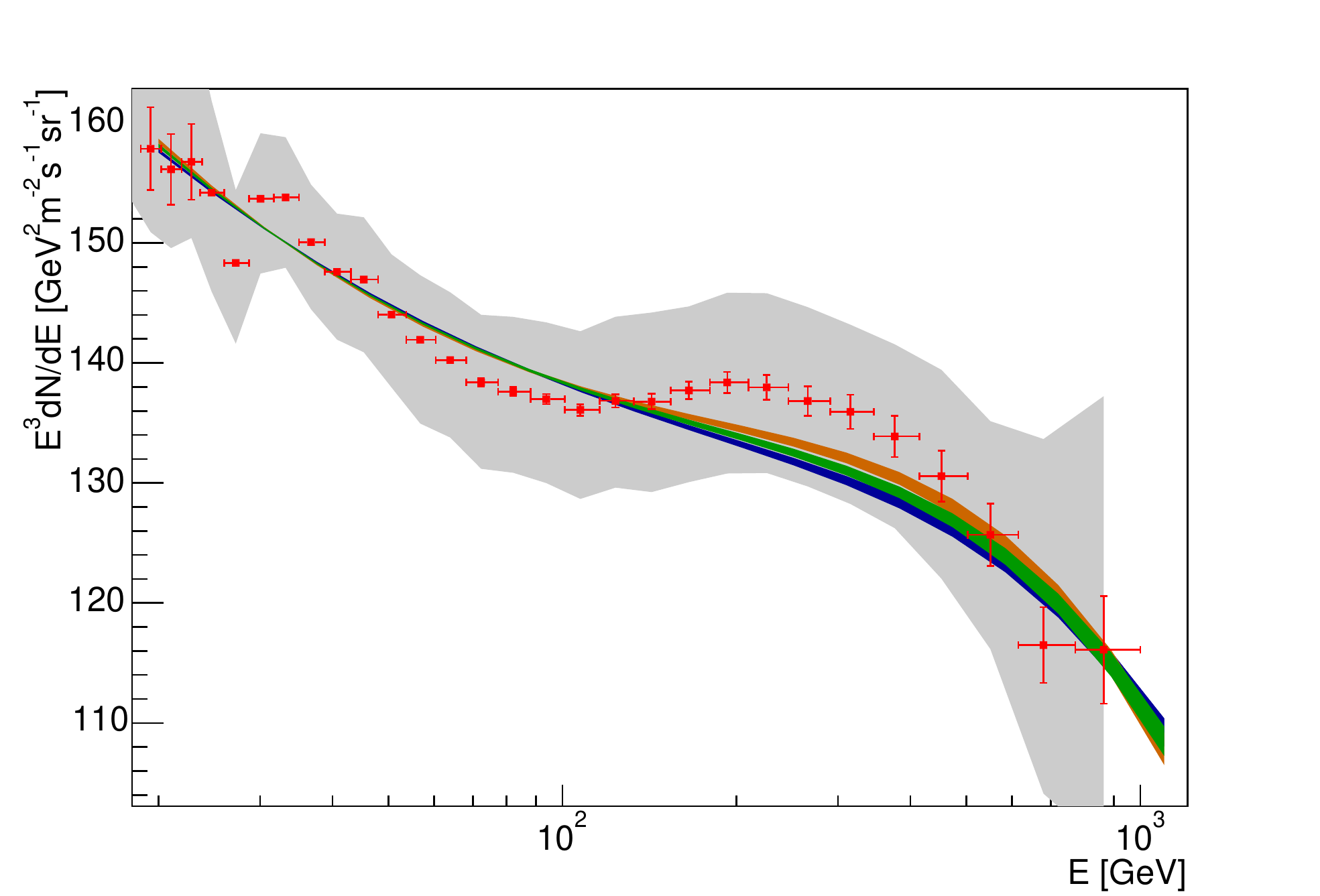}
\caption{CRE spectrum for the best fit model derived combining AMS-02 and Fermi-LAT data. Red points are Fermi-LAT experimental data. The blue area represents our best-fit model in the charge-symmetric scenario, with its 68\% uncertainty band. The green area is instead for the best fit model assuming charge asymmetry in the $\epsilon$-parameterization. The orange area is finally for the best fit model assuming charge asymmetry with the C component. Error bars are statistical errors, while systematic errors correspond to the grey band.}
\label{fig:bestfitcre}
\end{center}
\end{figure}

Concerning the combined AMS-02 and Fermi-LAT analysis, the situation is more complicated. Comparing Fig.~\ref{fig:AMSparams} and Fig.~\ref{fig:Fermiparams} it is clear that the maximum likelihood regions of AMS and Fermi-LAT do not show any overlap. The best-fit regions for AMS-02 correspond to a poor fit of Fermi-LAT data and vice-versa. The combined analysis (Fig.~\ref{fig:FermiAMSparams}) selects an intermediate region of the parameter space different from the individual regions preferred by Fermi-LAT and AMS-02, and the best-fit corresponds to the plot shown in Figures \ref{fig:bestfitpf} and \ref{fig:bestfitcre}. 

\begin{table}[tdp]
\caption{Best fit values and $1\sigma$ uncertainties for the parameters of our charge symmetric and charge asymmetric models. The model including component $C$ is not shown. Only statistical errors are accounted for. The parameter $E_{\rm cut,B}=1.6\pm0.1~\TeV$ in both cases. Reduced $\chi^{2}$ are shown without/with including Fermi-LAT systematics in the computation of the $\chi^{2}$.}
\begin{center}
\begin{tabular}{|c|c|c|c|c|c|c|c|}
\hline $A$ & $\alpha$ & $B$ & $\beta$ & $\epsilon$ & $\chi^{2}_{\rm AMS}$ & $\chi^{2}_{\rm Fermi}$ & $\chi^{2}_{\rm total}$\\
\hline $140\pm1$ & $3.166\pm0.003$ & $9.08\pm0.09$ & $2.61\pm0.01$ & 0 (fixed) & 1.3 & 9.7/0.22 & 4.7/0.76 \\
\hline $135\pm1$ & $3.193\pm0.004$ & $11.76\pm0.22$ & $2.63\pm0.01$  & $0.22\pm0.02$ & 1.18 & 9.1/0.18 & 4.2/0.64\\
\hline
\end{tabular}
\end{center}
\label{tab:fits}
\end{table}%

It is very important to point out now that only the statistical errors were taken into account in this analysis.  As it is shown in Tab.~\ref{tab:fits}, the reduced $\chi^{2}$~\footnote{ We report values of the $\chi^{2}$ only as an indicator of the goodness of fit. We remark that in our Bayesian approach $\chi^{2}$'s are never used to sample the parameter space. Rather, we use the likelihood estimator for that purpose.} corresponding to the Fermi-LAT fit in the best-fit model for the combined AMS-02 and Fermi-LAT datasets is of about 9, making it clear that Fermi-LAT and AMS-02 datasets are hardly compatible.  This is also confirmed by a Kolmogorov-Smirnov test on Fermi-LAT data \cite{NumericalRecipes,1992drea.book.....B}. This test is a very simple, yet powerful estimator of the difference between two cumulative distributions.\footnote{While the test is originally defined for unbinned distributions, it can also be used with binned ones.} In our case we test whether the distributions of Fermi-LAT data and of our best-fit model prediction are compatible. Compatibility is assessed in terms of a test statistics $D$ whose value depends on the maximum difference between the two cumulative distributions. Depending on the degrees of freedom of the problem, critical values of the test statistics are computed, corresponding to incompatibility at some confidence level if $D$ is larger than the related critical value. In our case, we have $D=0.90$ against a critical value of 0.29, meaning that Fermi-LAT data and our best-fit model are incompatible at more than 99.5\% confidence level. 

However, Fermi-LAT data are dominated by systematics in the whole energy range considered in this work, as it is clear from Fig.~\ref{fig:bestfitcre}. If we further account for systematic errors in Fermi-LAT measurements, the combined best-fit is not in strong tension with Fermi-LAT dataset ( at least if systematics are not correlated between energy bins), as clearly seen in Fig.~\ref{fig:bestfitcre} where the best-fit curve of the combined fit is demonstrated to be well within the systematics uncertainty band of Fermi-LAT. 
The reduced $\chi^2$ for the combined fit in this case is 0.76 (we compute it for illustrating purposes by adding systematic and statistical errors in quadrature). 

As a cross-check of this result, we run our MCMC procedure for the combined fit of AMS-02 and Fermi-LAT datasets in 3 different cases. In one case by adding a further parameter $F_{\rm norm}$ that shifts the overall normalization of Fermi-LAT points up and down and including this parameter in the combined fit. In the other two cases, by simply shifting the Fermi-LAT data points first up and then down by 1$\sigma_{\rm sys}$, thus making the hypothesis that systematics are perfectly correlated from one point to the next. We confirm that in all the 3 cases the best fit is only marginally affected. The best fit value of $F_{\rm norm}$ is 0.93 ($F_{\rm norm}=1$ means leaving the overall normalization of Fermi-LAT data points unchanged), implying a 7\% downward correction to Fermi-LAT data, of the same order as the systematic uncertainty, while the other parameters are only marginally affected. In particular the slopes $\alpha=3.169\pm0.003$ and $\beta=2.60\pm0.01$ are compatible within 1 standard deviation with the ones computed with statistical errors only and without allowing for a free normalization of Fermi-LAT data. The normalizations of the two components are more strongly affected on account of the different normalization of Fermi-LAT data.  Again, a Kolmogorov-Smirnov test on this model confirms that the fit of Fermi-LAT data is poor. This may then hint at the presence of energy-dependent effects in Fermi-LAT systematics. Similar results hold for the other two cases we considered. We can therefore conclude that the best fit model found in the combined case of AMS-02 and Fermi-LAT datasets is robust against possible large,  energy-dependent systematics errors in Fermi-LAT data and that the main effect of systematic uncertainties on our fits would be to increase the errors on best fit parameters. 

\begin{figure}[tbp]
\begin{center}
\includegraphics[width=\textwidth]{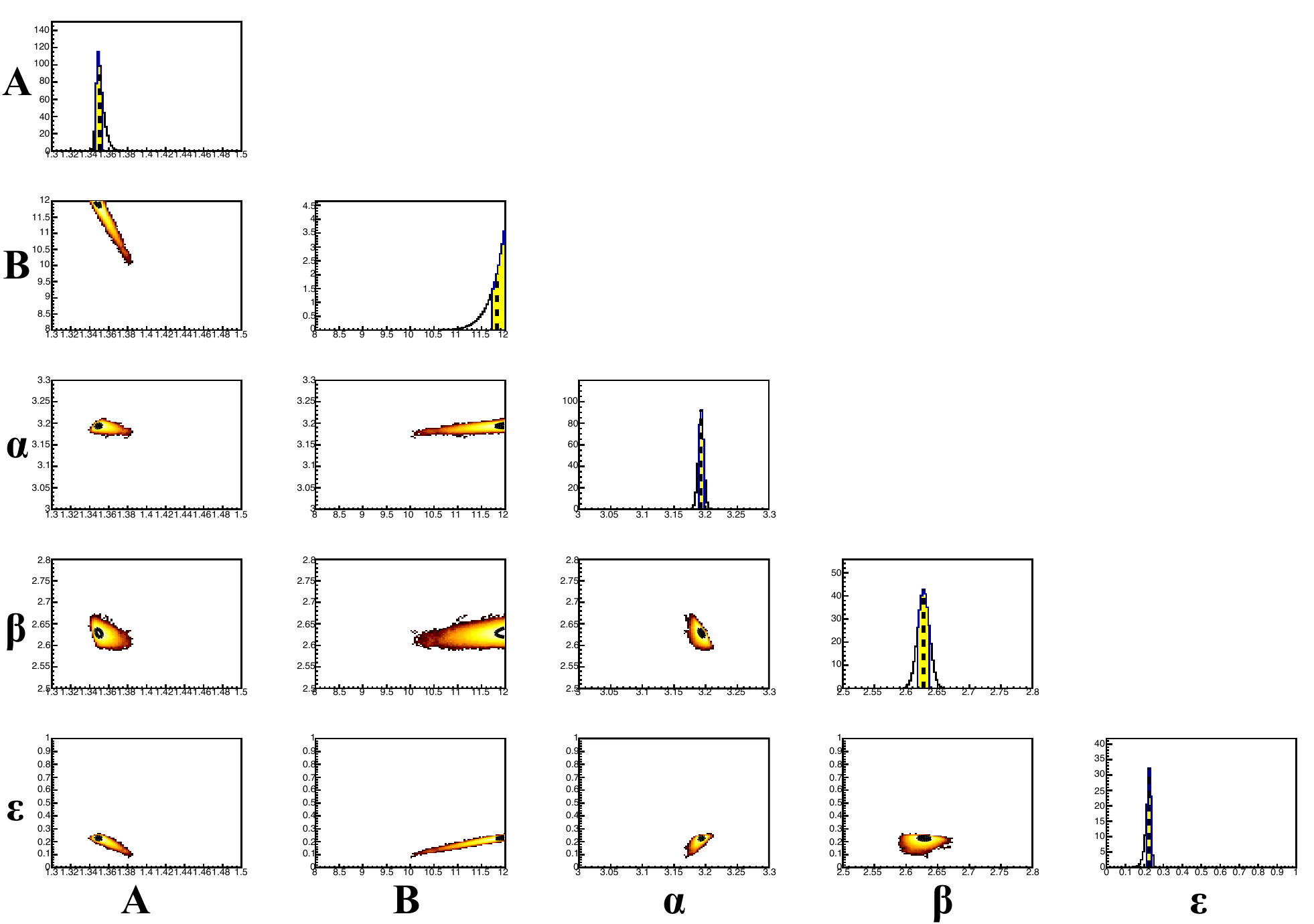}
\caption{Same format as Fig.~\ref{fig:AMSparams} for the combined AMS-02 and Fermi-LAT data in the case of charge asymmetry with $\epsilon$-parameterization. The distributions related to the parameter $E_{\rm cut,B}$ are not shown for plot clarity.}
\label{fig:FermiAMSasymmparams}
\end{center}
\end{figure}

\begin{figure}[tbp]
\begin{center}
\includegraphics[width=\textwidth]{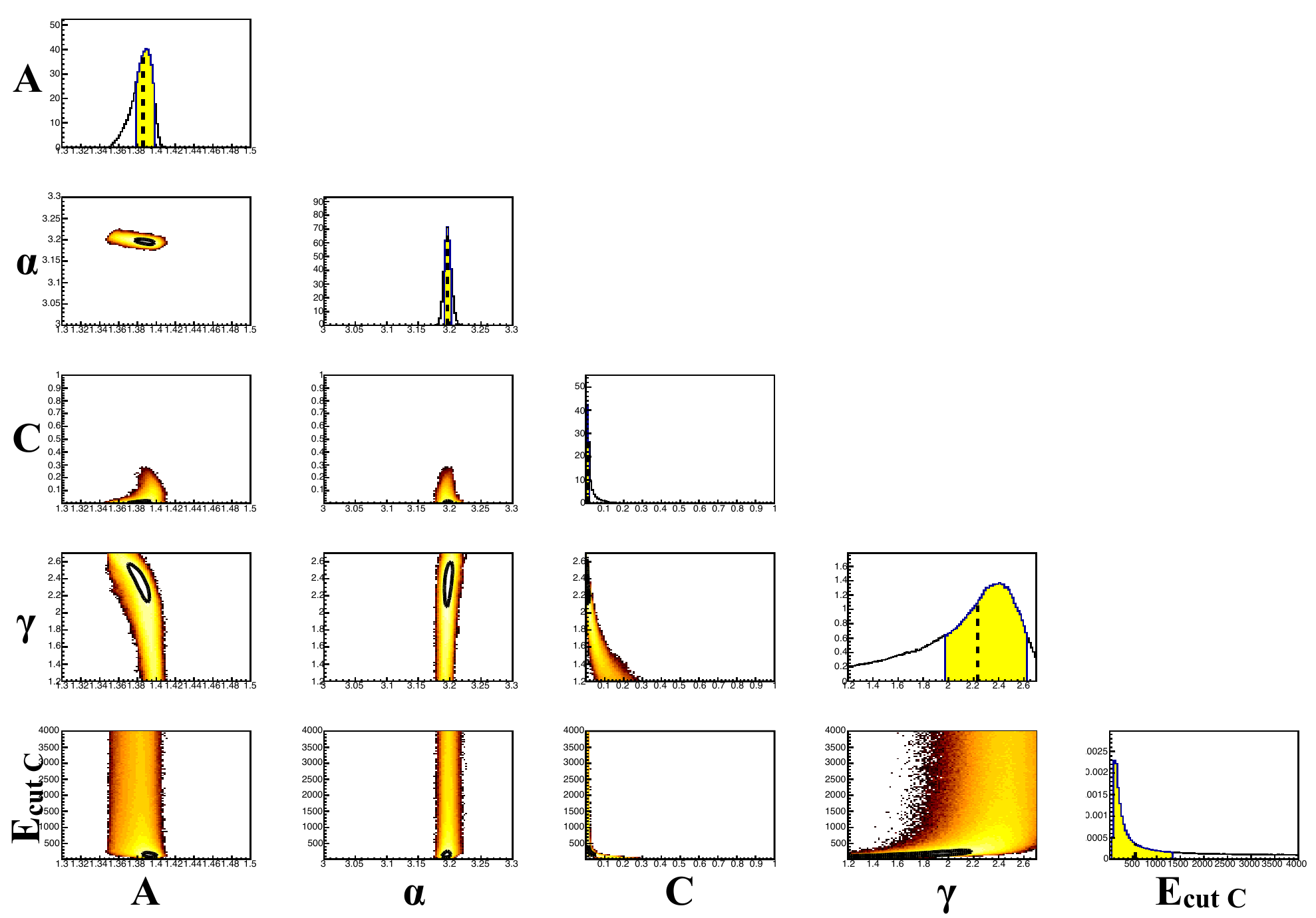}
\caption{Same format as Fig.~\ref{fig:AMSparams} for the combined AMS-02 and Fermi-LAT data in the case of charge asymmetry with the C component. For clarity the parameter distributions of component B are not shown.}
\label{fig:FermiAMSCparams}
\end{center}
\end{figure}

{ Now we consider deviations from charge-symmetric scenario}.
\begin{description}
\item[$\bm \epsilon$-parametrization] If we allow for an asymmetry in component $B$, with electrons corresponding to $(1+\epsilon)J_{B}$ and positrons to $(1-\epsilon)J_{B}$, we find a clear indication for $\epsilon\simeq0.22$. The spectral parameters of the component $B$ are not strongly affected (the parameters of the ``background'' component $A$ vary more strongly because of the additional electron contribution from the asymmetric $B$ component). The related results are shown in Fig.~\ref{fig:FermiAMSasymmparams}.~\footnote{The careful reader might notice that the marginalized distribution of parameter $B$ in this case is peaked towards the upper limit of the allowed range for this parameter. We prefer to show the marginalized distribution in this way for consistency with the corresponding plots for the other models. However, we verified that extending the range for parameter $B$ so that its probability distribution is fully sampled yields small adjustments of the best fit parameters and of the values of the $\chi^{2}$ but does not affect our conclusions.}
\item [C component] If we introduce a new population $C$ of electrons only, we find an indication of the possible presence of a very hard electronic component ($\gamma\simeq2$) which results to be active at relatively low energies, with $E_{\rm cut,C}\simeq300~\GeV$. 
Its normalization is however at the level of $10^{-3}$, about 5 orders of magnitude smaller than the other components, hinting at the fact that it might be only a mathematical artifact of the fitting procedure. { The related results shown in Fig.~\ref{fig:FermiAMSCparams} seem to confirm this impression. The sampling algorithm fails to converge to a definite region of the parameter space of component C and seems to prefer the lowest possible normalization. We remark that changing the parameter range or even the prior distribution (for example to a logarithmic prior for parameter $C$, so that the many orders of magnitude of its allowed range can be spanned more easily) does not affect our results.}
\end{description}

{ However, the improvement on the combined fit yielded by introducing either an asymmetry in $B$ or a new $C$ component is modest. In the case of $\epsilon$-parameterization, the} reduced $\chi^2$ of the combined AMS and Fermi-LAT fit improves from 0.76 to 0.64. An $F$-test \cite{1992drea.book.....B}, however, shows that this improvement is not sufficient to claim evidence for charge asymmetry at the 95\% confidence level (the same result holds if we consider the $\chi^{2}$ computed with statistical errors only). { We notice that} the improvement on the AMS-02 PF is intriguing: {reduced $\chi^2$ changes from 1.36 to 1.18 and the highest energy points are better reproduced. Future more accurate measurements of the positron fraction at higher energies can help distinguish the charge-symmetric from the charge-asymmetric scenarios (see Fig~\ref{fig:bestfitpf}).} 

{ In the C component case we obtain a similar result: the reduced $\chi^{2}$ for the combined analysis when C is included is 0.73, thus not improving much the fit over the charge symmetric model, in spite of the 3 additional free parameters.}

We show in Figures \ref{fig:bestfitpf} and \ref{fig:bestfitcre} a comparison between our best-fit models in the charge symmetric case and in the charge asymmetric ones for the PF and the CRE spectra respectively. The charge asymmetric model with $\epsilon$-parameterization slightly improves the agreement with the AMS-02 PF data at the highest energy points, but has only a very small impact on the agreement with Fermi-LAT data. { The addition of the C component instead yields a worsening of the PF, while not significantly improving on the agreement with Fermi-LAT data. Moreover, the large uncertainties in the parameter of the C component produces a very large uncertainty band in the PF above 200~GeV.}


\section{Conclusions}
\label{sec:conclusions}

{ We considered a model independent parametrization of the CR leptonic spectra as the superposition of a standard component  A plus an extra component B. A possible charge asymmetry in the extra component is described by either introducing a normalization offset between electron and positron spectra in component B ($\epsilon$-parameterization), or by leaving the B component charge symmetric and adding a further electron component C.} The spectra are modeled as power-laws in energy. We exploited Fermi-LAT, PAMELA and AMS-02 datasets and analyzed their mutual compatibility within this framework with a MCMC algorithm and using Bayesian inference. We report in Tab.~\ref{tab:fits} the best fit values of our parameters with their uncertainties.

{ In the pure charge symmetric scenario we find that the confidence regions in the parameter space for Fermi-LAT and AMS do not overlap, but the combined analysis produces a best-fit model which is good for AMS, and not in strong tension with Fermi-LAT, especially if we take into account systematic uncertainties.  Indeed, if Fermi-LAT systematic uncertainties are uncorrelated the goodness of the overall fit is very good. However, systematic errors are usually correlated. If they amount to an energy-independent overall rescaling the fit would not be good, but a mild energy-dependence to the systematics can bring the model and data into good agreement. Allowing for departures from charge symmetry in the extra component does not lead to substantial improvements to the fits. Therefore, the two datasets of AMS-02 and Fermi-LAT are compatible within systematic errors in a charge symmetric extra-component scenario and no compelling need for a charge asymmetry of the extra component emerges.}

\acknowledgments
We warmly thank Dario Grasso, Carmelo Evoli and Piero Ullio for reading the draft of this paper and providing useful insights.
DG thanks Piero Ullio and Wei Xue for inspiring discussions and the Max-Planck-Institute for Physics for warm hospitality during the initial phase of this work. 
LM acknowledges support from the Alexander von Humboldt foundation and partial support from the European Union FP7 ITN INVISIBLES (Marie Curie Actions, PITN- GA-2011- 289442).

\bibliographystyle{JHEP}
\bibliography{FitAMS02}

\providecommand{\href}[2]{#2}\begingroup\raggedright\begin{thebibliography}{10}

\bibitem{Aguilar:2013qda}
{\bf AMS Collaboration} Collaboration, M.~Aguilar et~al., {\it {First Result
  from the Alpha Magnetic Spectrometer on the International Space Station:
  Precision Measurement of the Positron Fraction in Primary Cosmic Rays of
  0.5–350 GeV}},  {\em Phys.Rev.Lett.} {\bf 110} (2013), no.~14 141102.

\bibitem{Adriani:2008zr}
{\bf PAMELA Collaboration} Collaboration, O.~Adriani et~al., {\it {An anomalous
  positron abundance in cosmic rays with energies 1.5-100 GeV}},  {\em Nature}
  {\bf 458} (2009) 607--609, [\href{http://xxx.lanl.gov/abs/0810.4995}{{\tt
  arXiv:0810.4995}}].

\bibitem{FermiLAT:2011ab}
{\bf Fermi LAT Collaboration} Collaboration, M.~Ackermann et~al., {\it
  {Measurement of separate cosmic-ray electron and positron spectra with the
  Fermi Large Area Telescope}},  {\em Phys.Rev.Lett.} {\bf 108} (2012) 011103,
  [\href{http://xxx.lanl.gov/abs/1109.0521}{{\tt arXiv:1109.0521}}].

\bibitem{Yuan:2013eba}
Q.~Yuan and X.-J. Bi, {\it {Reconcile the AMS-02 positron fraction and
  Fermi-LAT/HESS total $e^{\pm}$ spectra by the primary electron spectrum
  hardening}},  \href{http://xxx.lanl.gov/abs/1304.2687}{{\tt
  arXiv:1304.2687}}.

\bibitem{Yuan:2013eja}
Q.~Yuan, X.-J. Bi, G.-M. Chen, Y.-Q. Guo, S.-J. Lin, et~al., {\it {Implications
  of the AMS-02 positron fraction in cosmic rays}},
  \href{http://xxx.lanl.gov/abs/1304.1482}{{\tt arXiv:1304.1482}}.

\bibitem{Jin:2013nta}
H.-B. Jin, Y.-L. Wu, and Y.-F. Zhou, {\it {Implications of the first AMS-02
  measurement for dark matter annihilation and decay}},
  \href{http://xxx.lanl.gov/abs/1304.1997}{{\tt arXiv:1304.1997}}.

\bibitem{Cholis:2013psa}
I.~Cholis and D.~Hooper, {\it {Dark Matter and Pulsar Origins of the Rising
  Cosmic Ray Positron Fraction in Light of New Data From AMS}},
  \href{http://xxx.lanl.gov/abs/1304.1840}{{\tt arXiv:1304.1840}}.

\bibitem{Hooper:2008kg}
D.~Hooper, P.~Blasi, and P.~D. Serpico, {\it {Pulsars as the Sources of High
  Energy Cosmic Ray Positrons}},  {\em JCAP} {\bf 0901} (2009) 025,
  [\href{http://xxx.lanl.gov/abs/0810.1527}{{\tt arXiv:0810.1527}}].

\bibitem{Blasi:2009hv}
P.~Blasi, {\it {The origin of the positron excess in cosmic rays}},  {\em
  Phys.Rev.Lett.} {\bf 103} (2009) 051104,
  [\href{http://xxx.lanl.gov/abs/0903.2794}{{\tt arXiv:0903.2794}}].

\bibitem{Mertsch:2009ph}
P.~Mertsch and S.~Sarkar, {\it {Testing astrophysical models for the PAMELA
  positron excess with cosmic ray nuclei}},  {\em Phys.Rev.Lett.} {\bf 103}
  (2009) 081104, [\href{http://xxx.lanl.gov/abs/0905.3152}{{\tt
  arXiv:0905.3152}}].

\bibitem{Ahlers:2009ae}
M.~Ahlers, P.~Mertsch, and S.~Sarkar, {\it {On cosmic ray acceleration in
  supernova remnants and the FERMI/PAMELA data}},  {\em Phys.Rev.} {\bf D80}
  (2009) 123017, [\href{http://xxx.lanl.gov/abs/0909.4060}{{\tt
  arXiv:0909.4060}}].

\bibitem{Linden:2013mqa}
T.~Linden and S.~Profumo, {\it {Probing the Pulsar Origin of the Anomalous
  Positron Fraction with AMS-02 and Atmospheric Cherenkov Telescopes}},
  \href{http://xxx.lanl.gov/abs/1304.1791}{{\tt arXiv:1304.1791}}.

\bibitem{Abdo:2009zk}
{\bf Fermi LAT Collaboration} Collaboration, A.~A. Abdo et~al., {\it
  {Measurement of the Cosmic Ray e+ plus e- spectrum from 20 GeV to 1 TeV with
  the Fermi Large Area Telescope}},  {\em Phys.Rev.Lett.} {\bf 102} (2009)
  181101, [\href{http://xxx.lanl.gov/abs/0905.0025}{{\tt arXiv:0905.0025}}].

\bibitem{Ackermann:2010ij}
{\bf Fermi LAT Collaboration} Collaboration, M.~Ackermann et~al., {\it {Fermi
  LAT observations of cosmic-ray electrons from 7 GeV to 1 TeV}},  {\em
  Phys.Rev.} {\bf D82} (2010) 092004,
  [\href{http://xxx.lanl.gov/abs/1008.3999}{{\tt arXiv:1008.3999}}].

\bibitem{Masina:2013yea}
I.~Masina and F.~Sannino, {\it {Hints of a Charge Asymmetry in the Electron and
  Positron Cosmic-Ray Excesses}},
  \href{http://xxx.lanl.gov/abs/1304.2800}{{\tt arXiv:1304.2800}}.

\bibitem{Frandsen:2010mr}
M.~T. Frandsen, I.~Masina, and F.~Sannino, {\it {Cosmic Sum Rules}},  {\em
  Phys.Rev.} {\bf D83} (2011) 127301,
  [\href{http://xxx.lanl.gov/abs/1011.0013}{{\tt arXiv:1011.0013}}].

\bibitem{Masina:2011hu}
I.~Masina and F.~Sannino, {\it {Charge Asymmetric Cosmic Rays as a probe of
  Flavor Violating Asymmetric Dark Matter}},  {\em JCAP} {\bf 1109} (2011) 021,
  [\href{http://xxx.lanl.gov/abs/1106.3353}{{\tt arXiv:1106.3353}}].

\bibitem{Aharonian:2009ah}
{\bf H.E.S.S. Collaboration} Collaboration, F.~Aharonian et~al., {\it {Probing
  the ATIC peak in the cosmic-ray electron spectrum with H.E.S.S}},  {\em
  Astron.Astrophys.} {\bf 508} (2009) 561,
  [\href{http://xxx.lanl.gov/abs/0905.0105}{{\tt arXiv:0905.0105}}].

\bibitem{Adriani:2011xv}
{\bf PAMELA Collaboration} Collaboration, O.~Adriani et~al., {\it {The
  cosmic-ray electron flux measured by the PAMELA experiment between 1 and 625
  GeV}},  {\em Phys.Rev.Lett.} {\bf 106} (2011) 201101,
  [\href{http://xxx.lanl.gov/abs/1103.2880}{{\tt arXiv:1103.2880}}].

\bibitem{Berezinsky_book}
V.~S. {Berezinskii}, S.~V. {Bulanov}, V.~A. {Dogiel}, and V.~S. {Ptuskin}, {\em
  {Astrophysics of cosmic rays}}.
\newblock 1990.

\bibitem{Galpropweb}
``The galprop code for cosmic-ray transport and diffuse emission production.''
  \url{http://galprop.stanford.edu/}.

\bibitem{Strong:2007nh}
A.~W. Strong, I.~V. Moskalenko, and V.~S. Ptuskin, {\it {Cosmic-ray propagation
  and interactions in the Galaxy}},  {\em Ann.Rev.Nucl.Part.Sci.} {\bf 57}
  (2007) 285--327, [\href{http://xxx.lanl.gov/abs/astro-ph/0701517}{{\tt
  astro-ph/0701517}}].

\bibitem{Dragonweb}
``Dragon code.'' \url{http://www.dragonproject.org/}.

\bibitem{DiBernardo:2009ku}
G.~Di~Bernardo, C.~Evoli, D.~Gaggero, D.~Grasso, and L.~Maccione, {\it {Unified
  interpretation of cosmic-ray nuclei and antiproton recent measurements}},
  {\em Astropart.Phys.} {\bf 34} (2010) 274--283,
  [\href{http://xxx.lanl.gov/abs/0909.4548}{{\tt arXiv:0909.4548}}].

\bibitem{DiBernardo:2010is}
G.~Di~Bernardo, C.~Evoli, D.~Gaggero, D.~Grasso, L.~Maccione, et~al., {\it
  {Implications of the Cosmic Ray Electron Spectrum and Anisotropy measured
  with Fermi-LAT}},  {\em Astropart.Phys.} {\bf 34} (2011) 528--538,
  [\href{http://xxx.lanl.gov/abs/1010.0174}{{\tt arXiv:1010.0174}}].

\bibitem{Bulanov_1974Ap&SS}
S.~V. {Bulanov} and V.~A. {Dogel}, {\it {The Influence of the Energy Dependence
  of the Diffusion Coefficient on the Spectrum of the Electron Component of
  Cosmic Rays and the Radio Background Radiation of the Galaxy}},  {\em \apss}
  {\bf 29} (Aug., 1974) 305--318.

\bibitem{BATweb}
``Bat -- bayesian analysis toolkit.'' \url{http://www.mppmu.mpg.de/bat/}.

\bibitem{2009CoPhC.180.2197C}
A.~{Caldwell}, D.~{Koll{\'a}r}, and K.~{Kr{\"o}ninger}, {\it {BAT - The
  Bayesian analysis toolkit}},  {\em Computer Physics Communications} {\bf 180}
  (Nov., 2009) 2197--2209, [\href{http://xxx.lanl.gov/abs/0808.2552}{{\tt
  arXiv:0808.2552}}].

\bibitem{Malyshev:2009tw}
D.~Malyshev, I.~Cholis, and J.~Gelfand, {\it {Pulsars versus Dark Matter
  Interpretation of ATIC/PAMELA}},  {\em Phys.Rev.} {\bf D80} (2009) 063005,
  [\href{http://xxx.lanl.gov/abs/0903.1310}{{\tt arXiv:0903.1310}}].

\bibitem{NumericalRecipes}
W.~H. {Press}, S.~A. {Teukolsky}, W.~T. {Vetterling}, and B.~P. {Flannery},
  {\em {Numerical Recipes -- The Art of Scientific Computing}}.
\newblock Cambridge University Press, 2007.

\bibitem{1992drea.book.....B}
P.~R. {Bevington} and D.~K. {Robinson}, {\em {Data reduction and error analysis
  for the physical sciences}}.
\newblock 1992.

\end{thebibliography}\endgroup

%

\end{document}